\documentclass[prl,reprint,superscriptaddress,nofootinbib]{revtex4-1}  
\pdfoutput=1
\usepackage{amssymb,amsmath}
\usepackage{graphicx}
\usepackage{natbib}
\usepackage{mathrsfs}
\usepackage{color}
\usepackage[pdftex=true,linktoc=page,colorlinks,urlcolor=blue,citecolor=blue,linkcolor=blue]{hyperref}
\usepackage[letterpaper,textwidth=7in,top=.75in,bottom=.75in]{geometry}
\newcommand{\bra}[1]{\langle #1|}
\newcommand{\ket}[1]{|#1\rangle}

\begin{document}
\title{Electromagnetically-induced transparency in a diamond spin ensemble enables all-optical electromagnetic field sensing}

 \author{V.~M.~Acosta}
    \email{victor.acosta@hp.com}
    \altaffiliation{Equal contribution}
    \affiliation{
     Hewlett-Packard Laboratories, 1501 Page Mill Rd.,
     Palo Alto, CA 94304
    }

 \author{K.~Jensen}
    \email{kasperjensen@berkeley.edu}
    \altaffiliation{Equal contribution}
    \affiliation{
    Department of Physics, University of California-Berkeley,
    Berkeley CA 94720
    }

 \author{C.~Santori}
    \affiliation{
     Hewlett-Packard Laboratories, 1501 Page Mill Rd.,
     Palo Alto, CA 94304
    }
%

  \author{D.~Budker}
    \affiliation{
    Department of Physics, University of California-Berkeley,
    Berkeley CA 94720
    }

 \author{R.~G.~Beausoleil}
    \affiliation{
     Hewlett-Packard Laboratories, 1501 Page Mill Rd.,
     Palo Alto, CA 94304
    }

\begin{abstract}
We use electromagnetically-induced transparency (EIT) to probe the narrow electron-spin resonance of nitrogen-vacancy centers in diamond. Working with a multi-pass diamond chip at temperatures $6\mbox{-}30~{\rm K}$, the zero-phonon absorption line ($637~{\rm nm}$) exhibits an optical depth of $6$ and inhomogenous linewidth of ${\sim}30~{\rm GHz}$ full-width-at-half-maximum (FWHM). Simultaneous optical excitation at two frequencies separated by the ground-state zero-field splitting ($2.88~{\rm GHz}$), reveals EIT resonances with a contrast exceeding $6\%$ and FWHM down to $0.4~{\rm MHz}$. The resonances provide an all-optical probe of external electric and magnetic fields with a projected photon-shot-noise-limited sensitivity of $0.2~{\rm V/cm/\sqrt{Hz}}$ and $0.1~{\rm nT/\sqrt{Hz}}$, respectively. Operation of a prototype diamond-EIT magnetometer measures a noise floor of ${\lesssim}1~{\rm nT/\sqrt{Hz}}$ for frequencies above $10~{\rm Hz}$ and Allan deviation of $1.3{\pm}1.1~{\rm nT}$ for $100~{\rm s}$ intervals. The results demonstrate the potential of diamond-EIT devices for applications ranging from quantum-optical memory to precision measurement and tests of fundamental physics.

\end{abstract}
\maketitle
Electromagnetically-induced transparency (EIT) is an optical coherence effect which provides exquisite control over the absorption and dispersion in atomic media. In atoms with two coherent ground-state levels that can be optically excited to the same excited state (a ``$\Lambda$ system''), EIT results in ultra-narrow transmissive spectral features, with resonance quality factors exceeding $10^{12}$ \cite{BRA1997,BUD1999,KLE2006,ROH2012} and more than $10^7$ reduction in optical group velocity \cite{BUD1999,HAU1999}. Numerous EIT-based applications are being pursued, including precision measurement \cite{NAG1998,VAN2005,SAN2005,YUD2010}, few-photon nonlinear optics \cite{HAR1999,BAJ2009,ALB2011,PEY2012}, optical buffers \cite{KAS1995,KHU2005,BOY2005}, and quantum optical memories \cite{LVO2009,TIT2009,NOV2012}.

Critical to EIT-based applications is the simultaneous presence of substantial atomic absorption and long-lived ground-state coherence \cite{GOR2007}. This has motivated the use of atomic gases, where high optical depth (${\gg}1$) and long coherence times (${\gg}1~{\rm ms}$) can be simultaneously realized  \cite{LVO2009,NOV2012}. However, a solid-state approach is desirable for compatibility with large-scale fabrication processes. EIT in various rare-earth doped crystals has been observed \cite{HAM1997,GOL2009,BAL2010PRB}, and there has been substantial progress towards applications in quantum information. However, one drawback of many rare-earth-doped systems is weak optical transitions \cite{LVO2009,TIT2009}.

Ensembles of nitrogen-vacancy (NV) centers in diamond may provide an ideal compromise, owing to the relatively strong NV-light coupling \cite{DAV1976,BUC2010}, which can be enhanced using optical microcavities \cite{AHA2011,FAR2012,HAU2012,FAR2013}, and long ensemble spin coherence time (${\gg}100~{\rm ms}$ using decoupling techniques \cite{BAR2012}) at temperatures $T{\lesssim}100~{\rm K}$ \cite{FU2009,JAR2012}. EIT in diamond was observed before \cite{HEM2001}, but there the use of a high-defect-density diamond and the necessity of a large magnetic field (${\sim}0.1~{\rm T}$) limited the range of possible applications. More recent studies showed that $\Lambda$ systems can be realized near zero magnetic field \cite{SAN2006OPTEX,SAN2006,TAM2008,TOG2011,ACO2012}.

In this Letter, we report EIT with low-defect-density diamond at zero magnetic field. Using electron-irradiation and annealing techniques \cite{ACO2009} to enhance the NV absorption coefficient, $\alpha$, and a multi-pass diamond chip to increase the optical path length, $L$, we realize optical depth, $\alpha L>1$, and coherence time, $T_2^{\ast}>1~{\rm \mu s}$. EIT provides a means of probing the NV ground-state spin resonances without microwave irradiation, and we show that high-sensitivity, all-optical electric and magnetic field sensing is possible even in the presence of large bias electric fields. These are desirable features for cryogenic applications such as fundamental physics experiments \cite{ALT2009,BEC2011} and studies of novel superconductors \cite{BOU2011}. Our results are well-described by a model for EIT in inhomogenously-broadened media and can be extended to applications in nonlinear optics and quantum information.

\begin{figure}
\centering
    \includegraphics[width=.48\textwidth]{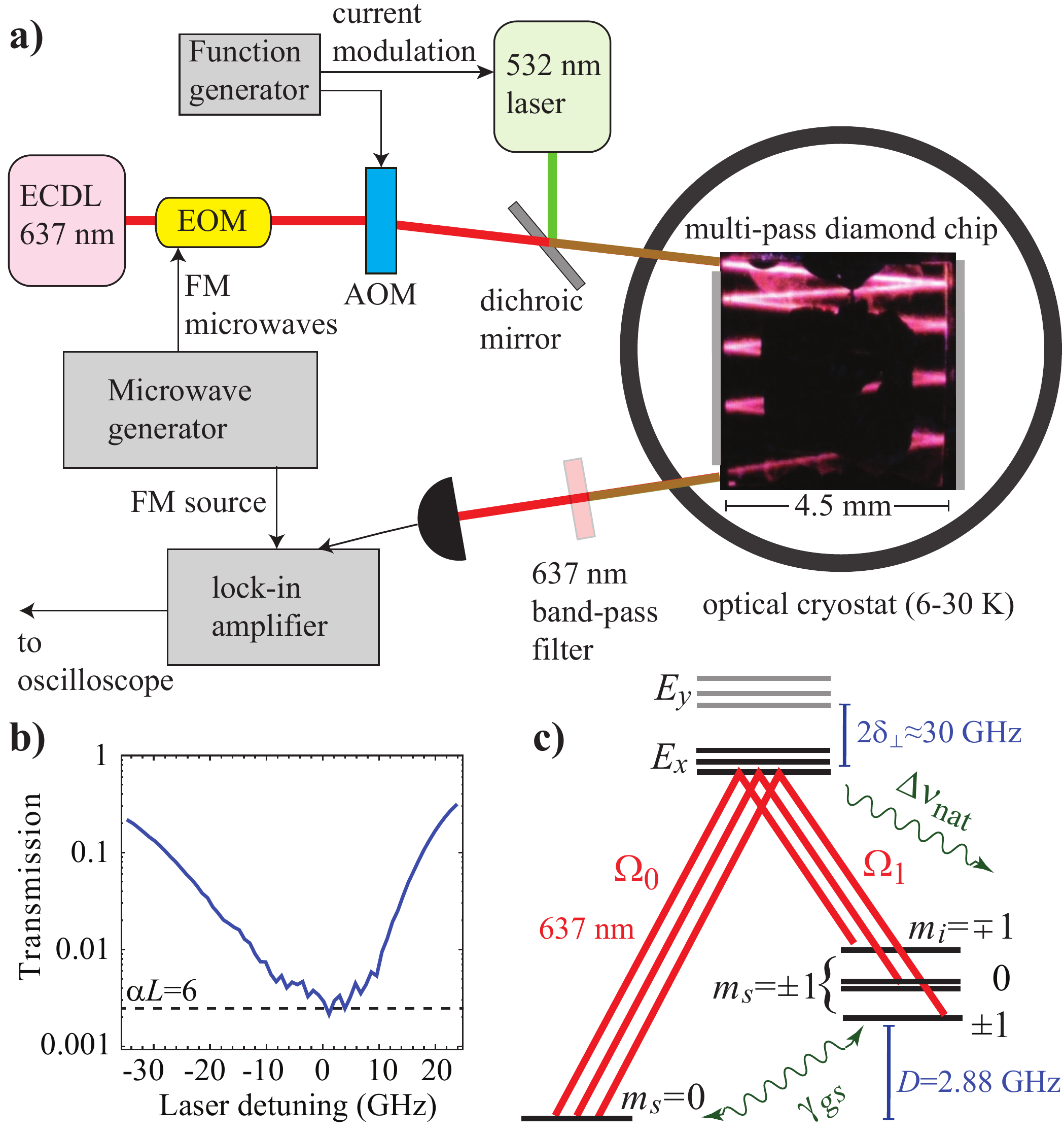}
    \caption{\label{fig:appEIT} (a) Experimental apparatus.  Microwave frequency modulation (FM) and lock-in detection were used for magnetometry (Fig.~\ref{fig:EITmag}). AOM--acousto-optic modulator. The dark square on the diamond surface is from a (disconnected) gold electrode. (b) Transmission spectrum of a weak optical probe.  (c) NV level structure and allowed $\Lambda$ transitions at $B=0$ in the moderate-strain regime. $^{14}$N quadrupole splitting is omitted.
    }
\end{figure}

The apparatus is depicted in Fig.~\ref{fig:appEIT}(a). Light $(637~{\rm nm})$ from an external-cavity diode laser (ECDL) was passed through an electro-optic phase modulator (EOM), producing sidebands with a tunable sideband-carrier detuning near $2.88~{\rm GHz}$. The EOM output was combined with a green repump beam ($532~{\rm nm}$), necessary to reverse optical bleaching \cite{HEM2001,SAN2006OPTEX,ASL2013}. The light beam then traveled 8 times through a multi-pass diamond chip housed in a continuous-flow liquid-helium cryostat. The transmitted light was spectrally filtered and detected by a photodiode. Experiments were performed at $T{\approx}10~{\rm K}$.

The diamond chip was a chemical-vapor-deposition-grown, single crystal with dimensions $4.5\times4.5\times0.5~{\rm mm}^3$ and nitrogen density $\rm [N]{\lesssim}1~ppm$. The sample was irradiated with $2~{\rm MeV}$ electrons (dose: $4\times10^{16}~{\rm cm^{-2}}$) and subsequently annealed at $800^{\circ}~{\rm C}$ for several hours. This resulted in $\rm [NV^{\mbox{-}}]{=}25{\pm}15~{\rm ppb}$, as measured by optical spectroscopy \cite{ACO2009}. Two opposing (100) sides of the diamond were polished and coated with Ag to produce 80-nm-thick mirrors. On one side, two windows (${\sim}0.5\times0.5~{\rm mm}^2$) were left uncoated to permit optical transmission.

This configuration allowed 8 passes through the diamond, limited by the angular deviation between the polished sides. An optical micrograph of the laser-induced fluorescence is shown within Fig.~\ref{fig:appEIT}(a). The beam (${\sim}100~{\rm \mu m}$ diameter) was collimated over the entire propagation length. Figure \ref{fig:appEIT}(b) shows the zero-phonon line transmission spectrum of a $500~{\rm nW}$ probe. The optical depth reaches $\alpha L{=}6$ with full-width-at-half-maximum (FWHM) ${\sim}30~{\rm GHz}$. Throughout, we normalize transmission by its off-resonant, room-temperature value, to account for interface losses, and the laser detuning, $\Delta_L$, is relative to absorption maximum ($470.480~{\rm THz}$).

Figure \ref{fig:appEIT}(c) illustrates the NV level structure. The center possesses $C_{3v}$ symmetry, with a paramagnetic ($S{=}1$) ground state and two $S{=}1$ excited-state orbitals \cite{MAZ2011,DOH2011}. Under moderate transverse strain, $\delta_{\perp}{=}15{\pm}10~{\rm GHz}$, level anticrossings in the lower excited-state orbital ($E_x$) mix electron spin projection, permitting optical transitions from both ground-state $m_s{=}0$ and $m_s{=}{\pm}1$ manifolds \cite{TAM2008,ACO2012}. The $m_s{=}0$ and $m_s{=}{\pm}1$ manifolds are split at zero magnetic field by $D{=}2.88~{\rm GHz}$, and hyperfine coupling with the $^{14}$N nucleus ($I{=}1$) results in three separate $\Lambda$ schemes.

\begin{figure}
\centering
    \includegraphics[width=.48\textwidth]{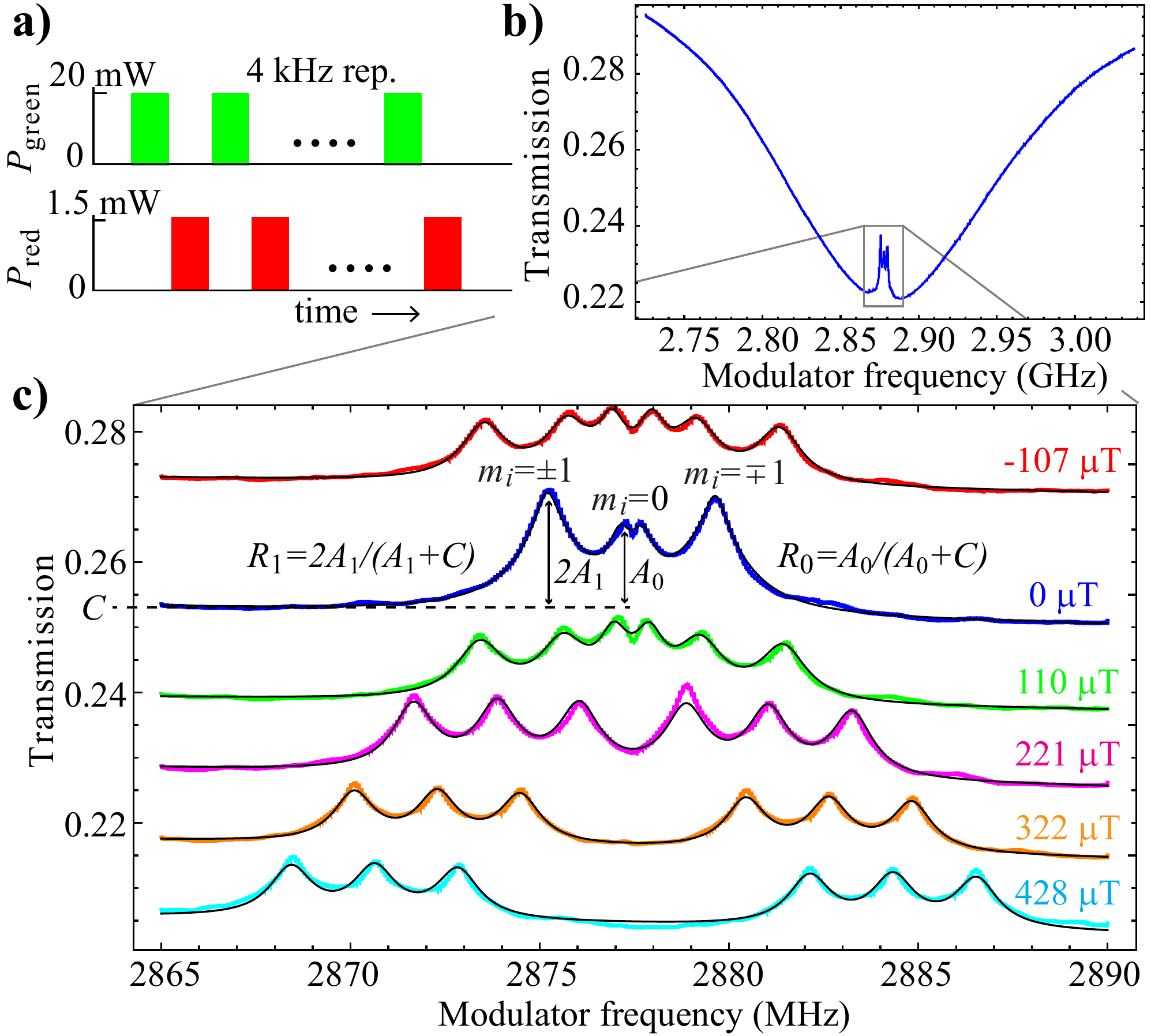}
    \caption{\label{fig:EITspectra} (a) Timing diagram of optical pulses used to probe EIT resonances. (b) Transmission spectrum as the sideband-carrier detuning was swept through two-photon resonance. (c) EIT spectra at different magnetic fields applied along a [100] direction. The spectra are offset for clarity. Overlayed are Lorentzian fits (see text). In (b,c) $\Delta_L{\approx}-15~{\rm GHz}$.
    }
\end{figure}

Figure \ref{fig:EITspectra}(a) shows the timing of optical pulses used to observe EIT. Alternating green and red pulses was necessary to simultaneously achieve high EIT contrast and minimize bleaching. To efficiently excite both arms of the $\Lambda$ transitions [Fig.~\ref{fig:appEIT}(c)], the sinusoidal optical phase modulation was set to yield a sideband:carrier:sideband intensity ratio of 0.7:1:0.7 (Supplementary Information, SI). Higher-order sidebands contributed $<8\%$ of the total intensity and are neglected in our analysis.

Figure \ref{fig:EITspectra}(b) shows the transmitted red light as a function of EOM drive frequency, $\nu_{\rm EOM}$. The broad anti-hole is due to optical pumping \cite{SAN2006OPTEX}. When $|\nu_{\rm EOM}-D|\gg\Delta\nu_{\rm nat}$, where $\Delta\nu_{\rm nat}{=}15~{\rm MHz}$ \cite{BAT2008} is the homogenous excited-state linewidth, NV centers resonant with one of the excitation frequencies can be excited from one spin sublevel, but are eventually trapped in the other sublevel, resulting in high transmission. As $\nu_{\rm EOM}$ approaches $2.88~{\rm GHz}$, one of the sidebands acts as a repump for the carrier, so there are no trap states, resulting in lower transmission. The FWHM of this feature (typically $50\mbox{-}200~{\rm MHz}$) depends on several factors, including excited-state dephasing \cite{FU2009}, optical power ($P_{\rm red}$), and spectral diffusion \cite{ACO2012,FAR2012}, but it is always $>\Delta\nu_{\rm nat}$.

As $\nu_{\rm EOM}$ matches exactly the ground-state splitting (two-photon resonance), the transmission increases sharply. These resonances are the hallmark of EIT, and they exhibit much narrower FWHM, $\Delta\nu_{\rm eit}{=}0.4\mbox{-}1.3~{\rm MHz}\ll\Delta\nu_{\rm nat}$. They occur because NV centers are optically pumped into a ``dark'' coherent superposition of ground-state levels, $\ket{D}$, which cannot interact with the light due to quantum interference. This can be understood by considering a simplified model for the NV center consisting of two ground-state levels, $m_s{=}1~(\ket{1})$ and $m_s{=}0~(\ket{0})$ driven optically to a single excited state, $\ket{E_x}$. The Hamiltonian under the rotating wave approximation is:
\begin{equation}
\begin{split}
\label{eq:ac}\mathscr{H}=&h/(4\pi)(\Omega_0\ket{E_x}\bra{0}+\Omega_1\ket{E_x}\bra{1})\\
&+h(\Delta_1{-}\Delta_0)\ket{1}\bra{1}+h\Delta_1\ket{E_x}\bra{E_x}+h.c.,
\end{split}
\end{equation}
where $h$ is Planck's constant, and $\Omega_{s}$ and $\Delta_{s}$ are, respectively, the Rabi frequency and detuning of the $\ket{s}\leftrightarrow\ket{E_x}$ transition. On two-photon resonance ($\Delta_1{=}\Delta_2$), the state, $\ket{D}{=}\frac{1}{\sqrt{\Omega_{0}^2+\Omega_{1}^2}}(\Omega_{1}\ket{0}{-}\Omega_{0}\ket{1})$, is completely decoupled from the optical fields, satisfying $\mathscr{H}\ket{D}{=}0$. The orthogonal superposition is coupled, so NV centers are pumped into $\ket{D}$, resulting in increased transmission.

The EIT linewidth is limited by the decoherence rate of the dark superposition, $\gamma_{gs}$, and its narrow width allows for sensitive, all-optical probing of the NV ground-state level structure. Figure \ref{fig:EITspectra}(c) shows EIT spectra for several values ($B$) of magnetic field along a [100] direction. [100]-directed fields preserve the degeneracy of the four NV axes, enabling higher EIT contrast. At $B{=}0$, the outermost resonances are split by ${\sim}2A_{HF}$, where $A_{HF}{=}-2.17~{\rm MHz}$ is the longitudinal hyperfine coupling constant \cite{FAN2013}. The small splitting $\delta_0{\approx}0.5~{\rm MHz}$ between the two innermost resonances ($m_i{=}0$) arises from transverse crystal strain, which behaves as an ensemble-averaged effective electric field \cite{HUG1967}. This effective electric field is $|E_{\perp}|{=}\delta_0/(2d_{gs_{\perp}}){\approx}15~{\rm kV/cm}$, where $d_{gs_{\perp}}{=}17~{\rm Hz/V/cm}$ is the ground-state transverse dipole moment \cite{VAN1990}.

The values of $\nu_{\rm EOM}$ on two-photon resonance are approximately equal to the ground-state transition frequencies \cite{DOL2011}:
\begin{equation}
 \label{eq:freqs}
 \nu_{{i}_{\pm}}{=}D\pm\sqrt{(g\mu_B B\cos{\theta}+m_iA_{HF})^2+(d_{gs_{\perp}}E_{\perp})^2},
\end{equation}
where $g{=}2.003$ is the electron-spin g-factor \cite{FEL2009}, $\mu_B{=}13.996~{\rm GHz/T}$ is the Bohr magneton, and $\cos{\theta}{\approx}1/\sqrt{3}$ is the projection of the field along each NV axis. Equation \eqref{eq:freqs} neglects AC Stark shifts of order $\Omega_0^2/(16\pi^2\Delta_{e}){\lesssim}10~{\rm kHz}$, where $\Delta_e{\approx}0.5~{\rm GHz}$ is a typical frequency separation between excited states (SI).

We fit the spectra to a sum of six Lorentzian profiles (plus an offset $C$), with amplitude $A$, and FWHM $\Delta\nu_{\rm eit}$. The central frequencies were constrained by Eq.~\eqref{eq:freqs}; $D$ and $E_{\perp}$ were fit globally, while $B$, $\Delta\nu_{\rm eit}$, $A$, and $C$ were allowed to vary between spectra. Resonances that satisfied $|g\mu_B B\cos{\theta}+m_iA_{HF}|<d_{gs_{\perp}}|E_{\perp}|$ (innermost resonances within the top three spectra in Fig.~\ref{fig:EITspectra}) were given separate amplitude ($A_0$) and FWHM ($\Delta\nu_{\rm eit,0}$) from those where the condition was not held ($A_1$, $\Delta\nu_{\rm eit,1}$) \cite{DOL2011}.


\begin{figure}
\centering
    \includegraphics[width=.48\textwidth]{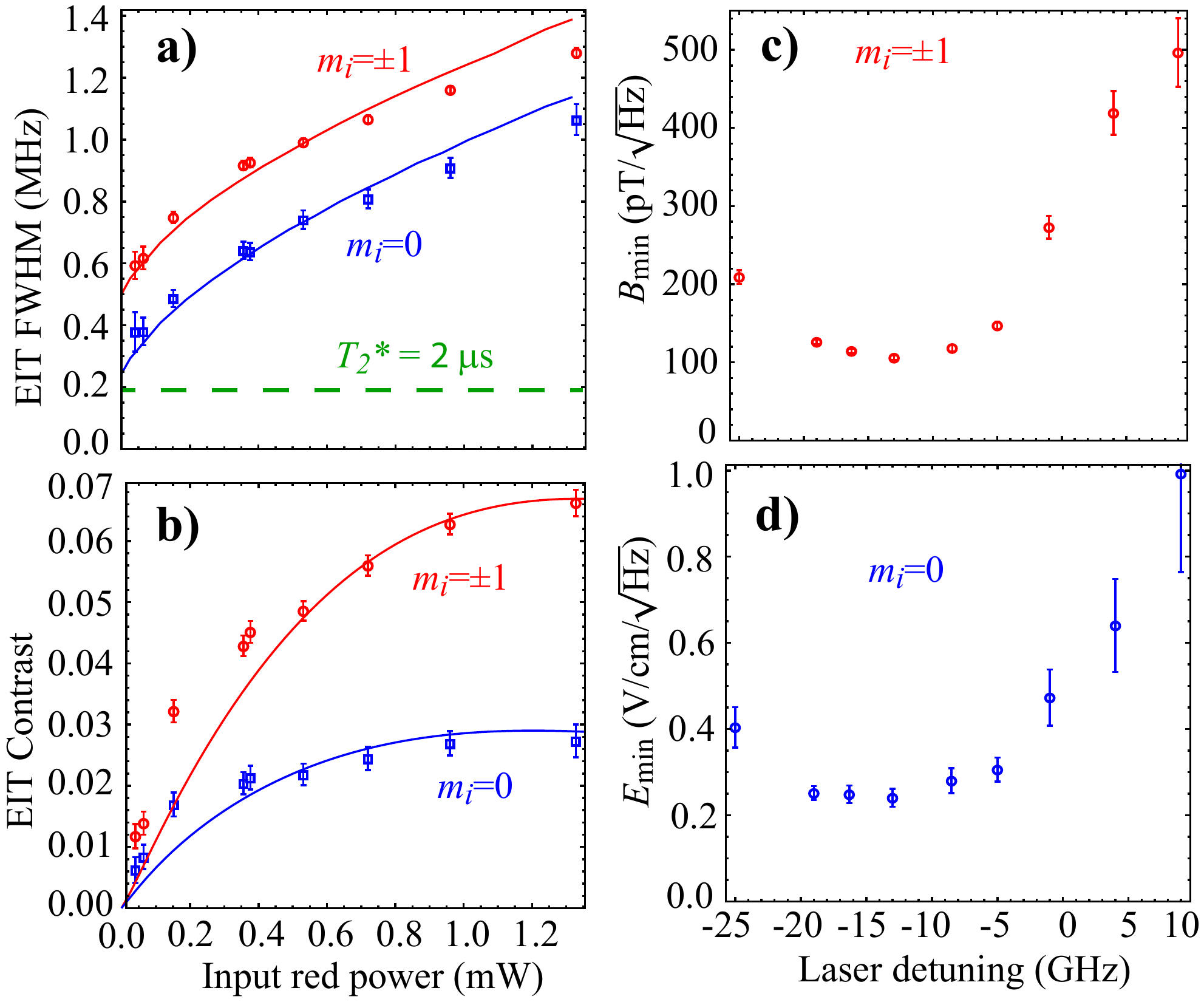}
    \caption{\label{fig:parameters} (a,b) Power dependence of zero-field EIT FWHM and contrast ($\Delta_L{\approx}-13~{\rm GHz}$) determined from Lorentzian fits as in Fig.~\ref{fig:EITspectra}(c). (c,d) Photon-shot-noise-limited magnetic and electric field sensitivity [Eq.~\eqref{eq:Bmin}], inferred from zero-field EIT spectra, as a function of laser detuning.
    }
\end{figure}

To gain further insight into the EIT lineshapes, we determined the contrast and FWHM as a function of $P_{\rm red}$ at $B{=}0$, Fig.~\ref{fig:parameters}(a,b). For $m_i{=}0$ resonances, the zero-field contrast is here defined as $R_0{=}A_0/(A_0+C)$, and for $m_i{=}{\pm}1$ resonances it is $R_1{=}2A_1/(A_1+C)$.  These data were fit by a 3-level density-matrix model for EIT in inhomogenously-broadened media (SI) \cite{SAN2006OPTEX}. The fit parameters include a Rabi frequency conversion factor, $P_{\rm sat}{\equiv}\pi\Delta\nu_{\rm nat}^{2}P_{\rm red}/\Omega_{0}^2{=}2.4{\pm}1.1~{\rm mW}$, the ratio $\Omega_{1}/\Omega_{0}{=}0.08{\pm}0.02$, which reflects the ensemble-averaged asymmetry in $\Lambda$ transition strengths \cite{ACO2012}, and nuclear-spin-dependent ground-state dephasing \cite{DOL2011} $\gamma_{gs,m_i{=}{\pm}1}/(2\pi){=}240{\pm}92~{\rm kHz}$ and $\gamma_{gs,m_i{=}0}/(2\pi){=}99{\pm}30~{\rm kHz}$. The apparent saturation of the contrast arises from photo-ionization (SI), which reduces $\rm [NV^{\mbox{-}}]$ at a rate ${\propto}P_{\rm red}^2$ \cite{ASL2013}. Above $T{\approx}10~{\rm K}$, the contrast falls off sharply with temperature, becoming negligible at $T{\approx}30~{\rm K}$. This effect (SI) is well described by a nine-level model of the NV center that includes temperature-dependent excited-state dephasing \cite{FU2009}.



Near $B{\approx}0$, the EIT resonances provide a means to simultaneously sense electric and magnetic fields. The theoretical sensitivity of an electric or magnetic field sensor based on an optically-detected signal, $S$, is given by the minimum detectable field which gives signal:noise${=}1$; $\delta E_{min}{=}\frac{\delta S}{|dS/dE|}$ and $\delta B_{min}{=}\frac{\delta S}{|dS/dB|}$, where $\delta S$ is the standard deviation of $S$ \cite{BUD2007}. From Eq.~\eqref{eq:freqs}, we see that $dS/dE{\approx}0$ except when $|g\mu_B B+m_i A_{HF}|{\lesssim}d_{gs_{\perp}}|E_{\perp}|$. At $B{\approx}0$, this condition is satisfied for the $m_i{=}0$ resonances, whereas high-sensitivity magnetometry is possible using the $m_i{=}{\pm}1$ resonances provided $|E_{\perp}|{\lesssim}|A_{HF}|/d_{gs_{\perp}}{\approx}130~{\rm kV/cm}$.

If $\delta S$ is limited by photon shot noise, the magnetic sensitivity is (SI):
\begin{equation}
\label{eq:Bmin}
\delta B_{min} \simeq \frac{1}{g\mu_B \cos{\theta}}\frac{\Delta\nu_{\rm eit}}{R_1} \sqrt{\frac{E_p}{P t_m}},
\end{equation}
where $P$ is the detected optical power, $E_p$ is the photon energy, and $t_m$ is the measurement time. A similar expression can be found for sensitivity to electric fields within diamond (SI), substituting $d_{gs_{\perp}}R_0$ for $g\mu_B R_1$ \cite{DOL2011} (neglecting crystal-strain inhomogeneity). These equations assume detected fields lie along a [100] direction, but arbitrarily-oriented fields can be detected using a suitable bias field \cite{TAY2008,MAE2010}.


We studied the projected electrometer and magnetometer performance by analyzing zero-field EIT spectra as a function of $\Delta_L$, Fig.~\ref{fig:parameters}(c-d). In both cases, the sensitivity is optimized in the range $\Delta_L{=}-5$ to $-20~{\rm GHz}$. Not surprisingly, this corresponds to the range of $\ket{E_x}$ strain shifts where $\Lambda$ transitions have been observed in single-NV experiments \cite{ACO2012}. The optimal sensitivities ($\Delta_L{=}-13~{\rm GHz}$) are $105{\pm}6~{\rm pT/\sqrt{Hz}}$ and $0.24{\pm}0.04~{\rm V/cm/\sqrt{Hz}}$ for magnetometry and electrometry, respectively. Throughout we use the metrology convention that 1 Hz measurement bandwidth corresponds to $t_m=0.5~{\rm s}$.

\begin{figure}
\centering
    \includegraphics[width=.48\textwidth]{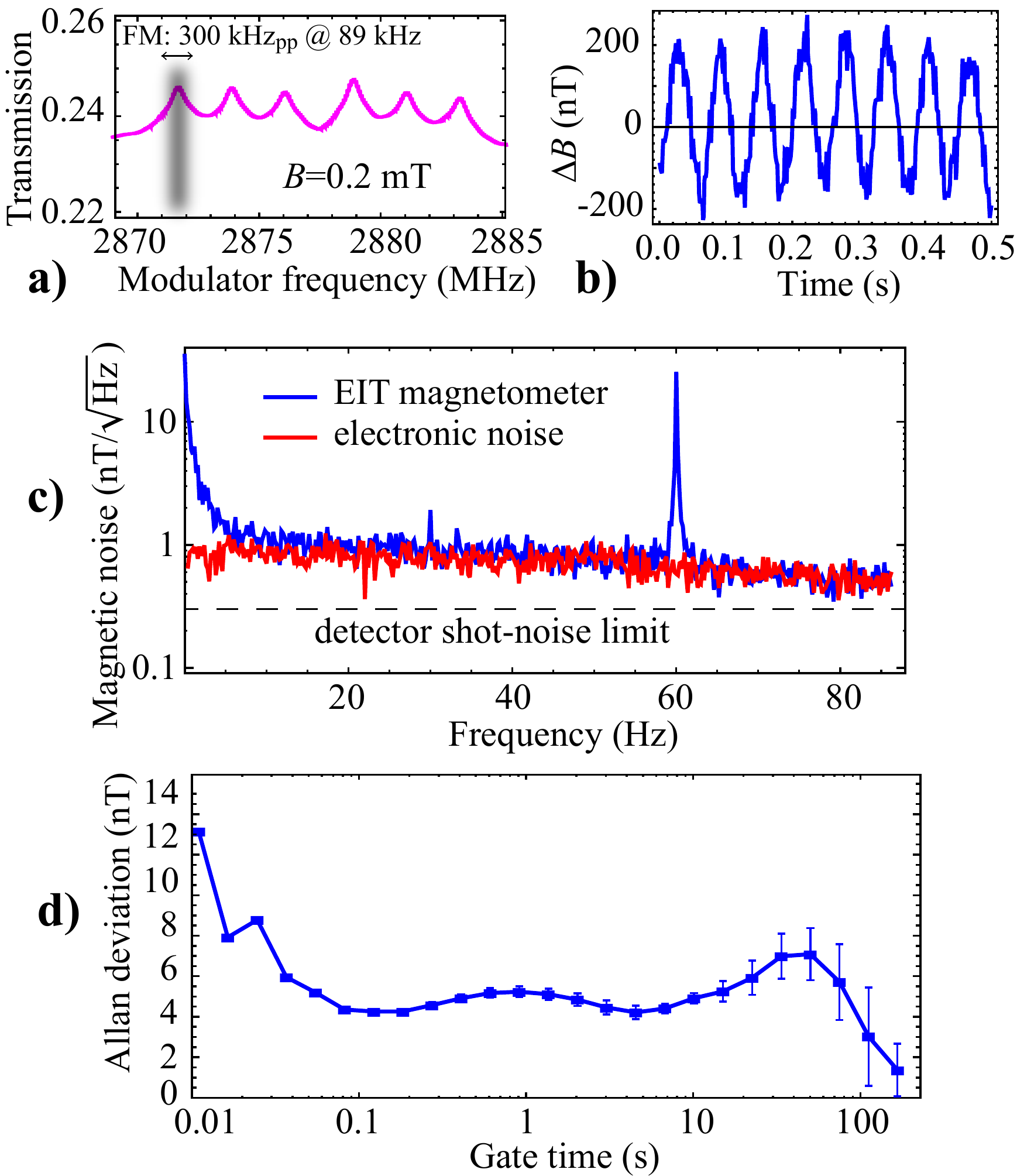}
    \caption{\label{fig:EITmag} (a) Schematic of the frequency modulation (FM) technique employed for magnetometry.  (b) Lock-in signal when a $16$-Hz oscillating field ($350~{\rm nT_{pp}}$) was applied along a [100] direction. (c) Magnetic noise spectrum of the EIT magnetometer. The spike at 60 Hz is a real magnetic signal arising due to operation in an unshielded environment (SI). Also shown is the noise spectrum when the photodetector was unplugged. (d) Allan deviation of a $500~{\rm s}$ data set.
    }
\end{figure}

High-sensitivity operation of our device as an all-optical magnetometer was accomplished using lock-in detection of a resonance peak \cite{ACO2010APL} in a bias field $B{=}0.2~{\rm mT}$ [Fig.~\ref{fig:EITmag}(a)]. A microwave signal $\nu_{\rm EOM}{=}2.8717~{\rm GHz}$ was frequency-modulated at $89~{\rm kHz}$ \cite{SHI2012}, with deviation $300~{\rm kHz_{pp}}$, and sent to the EOM. The photodetector signal was demodulated using a lock-in amplifier, and the in-phase signal was sent to an oscilloscope. Figure \ref{fig:EITmag}(b) shows the magnetometer response when a separately-calibrated test field was applied. Additional calibrations are presented in SI. Summarizing, the magnetometer response remains linear over a range of $\gtrsim1~{\rm \mu T}$, and the bandwidth covers $\sim100~{\rm Hz}$ (limited here by electronic filtering).

To characterize the sensitivity, the magnetic noise was measured without a test field. Figure \ref{fig:EITmag}(c) shows the noise-equivalent magnetic field spectrum.  The noise floor is ${\lesssim}1~{\rm nT/\sqrt{Hz}}$ for frequencies above $10~{\rm Hz}$. The floor was dominated by lock-in amplifier input noise, as evidenced by its persistence when the photodetector was unplugged. The expected shot noise, based on Eq.~\eqref{eq:Bmin} after incorporating the finite quantum efficiency of the detector, is $0.3~{\rm nT/\sqrt{Hz}}$.

Near $1~{\rm Hz}$, the noise floor rises. A possible cause is instabilities in $P_{\rm red}$ lead to fluctuations in AC Stark shifts (SI). Nevertheless, the magnetometer recovers high sensitivity for long integration times, as evidenced by the Allan deviation \cite{FAN2013} plotted in Fig.~\ref{fig:EITmag}(d). For a gate time of $100~{\rm s}$, the Allan deviation is $1.3{\pm}1.1~{\rm nT}$.

In comparing with existing technologies, we are not aware of another sensor that simultaneously measures low-frequency electric and magnetic fields with high sensitivity. Miniature vapor-cell magnetometers \cite{GRI2010} achieve low-frequency sensitivity of ${\sim}0.01~{\rm pT/\sqrt{Hz}}$ at $T{\approx}470~{\rm K}$, but the sensitivity rapidly degrades with decreasing temperature \cite{BUD2007}. At low temperature, miniature superconducting quantum interference devices have excellent sensitivity, $\delta B_{min}{\lesssim}0.1~{\rm pT/\sqrt{Hz}}$. However they typically suffer from $1/f$ noise, so DC measurements require external calibration \cite{CLA2006}. In addition to its long-term stability and dual electric/magnetic field sensitivity, our sensor is probed entirely optically, in large electric fields, and it does not produce fields of its own.

Future applications may benefit from several improvements. Higher EIT contrast is possible using spectral filtering to detect only one optical frequency, tailoring the absorption spectrum using holeburning techniques \cite{RED1987}, or employing other magnetic/electric field geometries \cite{HEM2001,ACO2012}. The coherence time can be extended by orders of magnitude using dynamic decoupling \cite{BAR2012} and/or with isotopically-pure diamond \cite{BAL2009,ISH2012,FAN2013}. Finally, integration with optical cavities \cite{AHA2011,HAU2012,FAR2013} will allow higher optical depth and larger optical intensities, while decreasing the device volume.

In summary, we observed narrow EIT resonances in a multi-pass diamond chip. The high optical depth and narrow inhomogenous linewidth enable new diamond-based applications including all-optical electrometry and magnetometry. We operated a prototype diamond-EIT magnetometer with sub-${\rm nT/\sqrt{Hz}}$ sensitivity and excellent long-term stability. Integration with photonic networks may enable new applications, including few-photon nonlinear optics and quantum-optical memories.

We thank P. Hemmer, D. Beck, Z. Huang, K. Heshami, P. Barclay, C. Simon, and A. Faraon for fruitful discussions. We acknowledge support by the Defense Advanced Research Projects Agency (award no. HR0011-09-1-0006) and the Regents of the University of California. D. B. acknowledges support from AFOSR/DARPA QuASAR program, NSF, and IMOD. K. J. was supported by the Danish Council for Independent Research | Natural Sciences.

\clearpage

\begin{widetext}
\begin{center}
\textbf{\large Supplementary Information: Electromagnetically-induced transparency in a diamond spin ensemble enables all-optical electromagnetic field sensing}
\end{center}
\end{widetext}

\setcounter{figure}{0}
\setcounter{equation}{0}
\setcounter{page}{1}
\makeatletter
\renewcommand{\thefigure}{S\@arabic\c@figure}
\renewcommand{\theequation}{S\@arabic\c@equation}
\renewcommand{\bibnumfmt}[1]{[S#1]}
\renewcommand{\citenumfont}[1]{S#1}

\section{Sideband power optimization}
\begin{figure}
\centering
    \includegraphics[width=.4\textwidth]{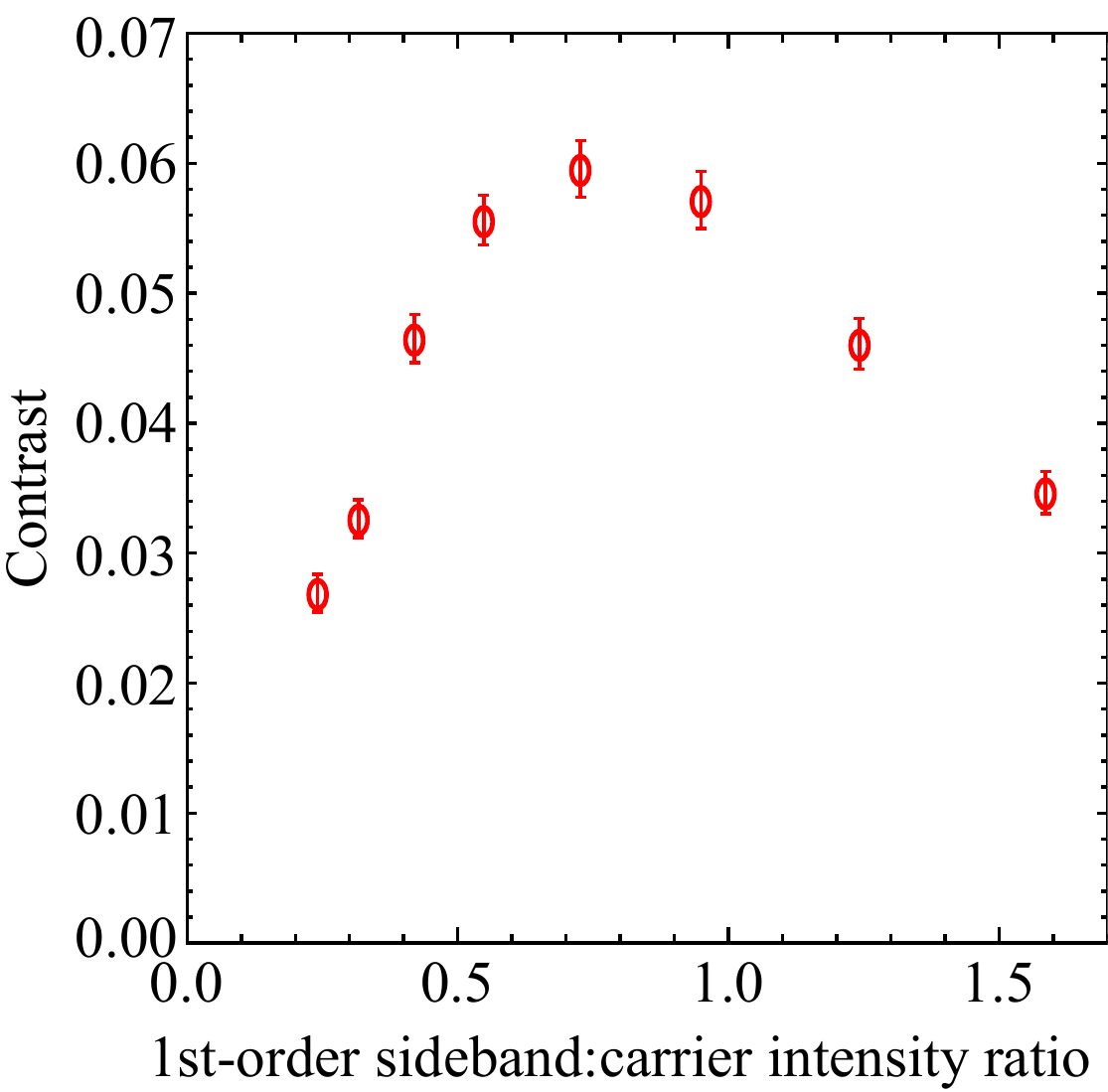}
    \caption{\label{fig:sideband} Contrast of zero-field $m_i{=}{\pm}1$ EIT resonances as a function of sideband:carrier intensity ratio.}
\end{figure}

In order to determine the optimal sideband:carrier intensity ratio, we varied the microwave power driving our EOM and fit the $B{=}0$ EIT spectra to obtain the contrast. Figure \ref{fig:sideband} shows the EIT contrast versus the first-order sideband:carrier intensity ratio, at constant $P_{\rm red}{=}1.3~{\rm mW}$. The sideband:carrier ratios were determined using a scanning Fabry-Perot interferometer. For ratios below $1$, the higher-order sidebands contribute ${\lesssim}10\%$ to the total intensity and can be neglected. For ratios above 1, the higher-order sidebands become substantial and lead to a reduction in contrast, as the intensity for any given frequency is lower. In the experiments presented in the main text, we chose a sideband:carrier:sideband intensity ratio of 0.7:1:0.7, where the maximum contrast was observed.

\section{Modeling EIT resonances at \emph{T}$\lesssim$10 K: 3-level model}
At temperature $T{\lesssim}10~{\rm K}$, the excited-state levels of the NV center are well resolved \cite{FU2009S}, with typical frequency spacing, $\Delta_e{\approx}0.5~{\rm GHz}$. Consequently, for relatively weak optical excitation, $\Omega_{0,1}<<2\pi\Delta_e$, we can treat our system as an ensemble of three-level atoms. The situation at higher temperature, where phonon interactions lead to significant broadening of the excited states, is treated in the next section.

We begin by writing the rotating-wave Hamiltonian, as in Eq.~\eqref{eq:ac} in the main text, but now in matrix form:
\begin{equation}
\label{eq:H}
\mathscr{H}/\hbar=\left(
\begin{array}{cccc}
0 & 0 & \Omega _{0}/2 \\
0 & 2\pi\delta & \Omega _{1}/2 \\
\Omega _{0}/2 & \Omega _{1}/2 & 2\pi\Delta_{0}
\end{array}
\right),
\end{equation}
where the basis is $\{\ket{0},\ket{1},\ket{E_x}\}$. Here $\Delta_s$ is the detuning from the $\ket{s}\leftrightarrow\ket{E_x}$ resonance (``one-photon detuning''), $\delta{=}\Delta_0-\Delta_1$ is the detuning from two-photon resonance, and we have assumed real Rabi frequencies $\Omega_{s}^{\ast}{=}\Omega_{s}$. Note that at zero field, the $m_s{=}{\pm}1$ resonances are degenerate, so in principle the system involves four levels. However only one superposition of $m_s{=}{\pm}1$ can be optically coupled to $\ket{E_x}$, so we label this superposition $\ket{1}$ and ignore the uncoupled, orthogonal superposition. Unlike the EIT dark state, defined in the main text, this uncoupled superposition does not play any role in the EIT spectrum, as it does not include $\ket{0}$.

In order to compute steady-state solutions, we use a master equation for the density matrix, $\rho$, in the presence of relaxation, $\mathscr{R}[\rho]$:
\begin{equation}
\label{eq:master}
d\rho/dt=-i[\mathscr{H}/\hbar,\rho]+\mathscr{R}[\rho]=0.
\end{equation}
Here $\mathscr{R}$ includes spontaneous emission, $\Delta\nu_{\rm nat}{=}15~{\rm MHz}$, as well as nuclear-spin-dependent ground-state decoherence, $\gamma_{gs,m_i}$, and longitudinal ground-state relaxation, $\gamma_1$. We set all other dephasing terms to zero.

\begin{figure}
\centering
    \includegraphics[width=.4\textwidth]{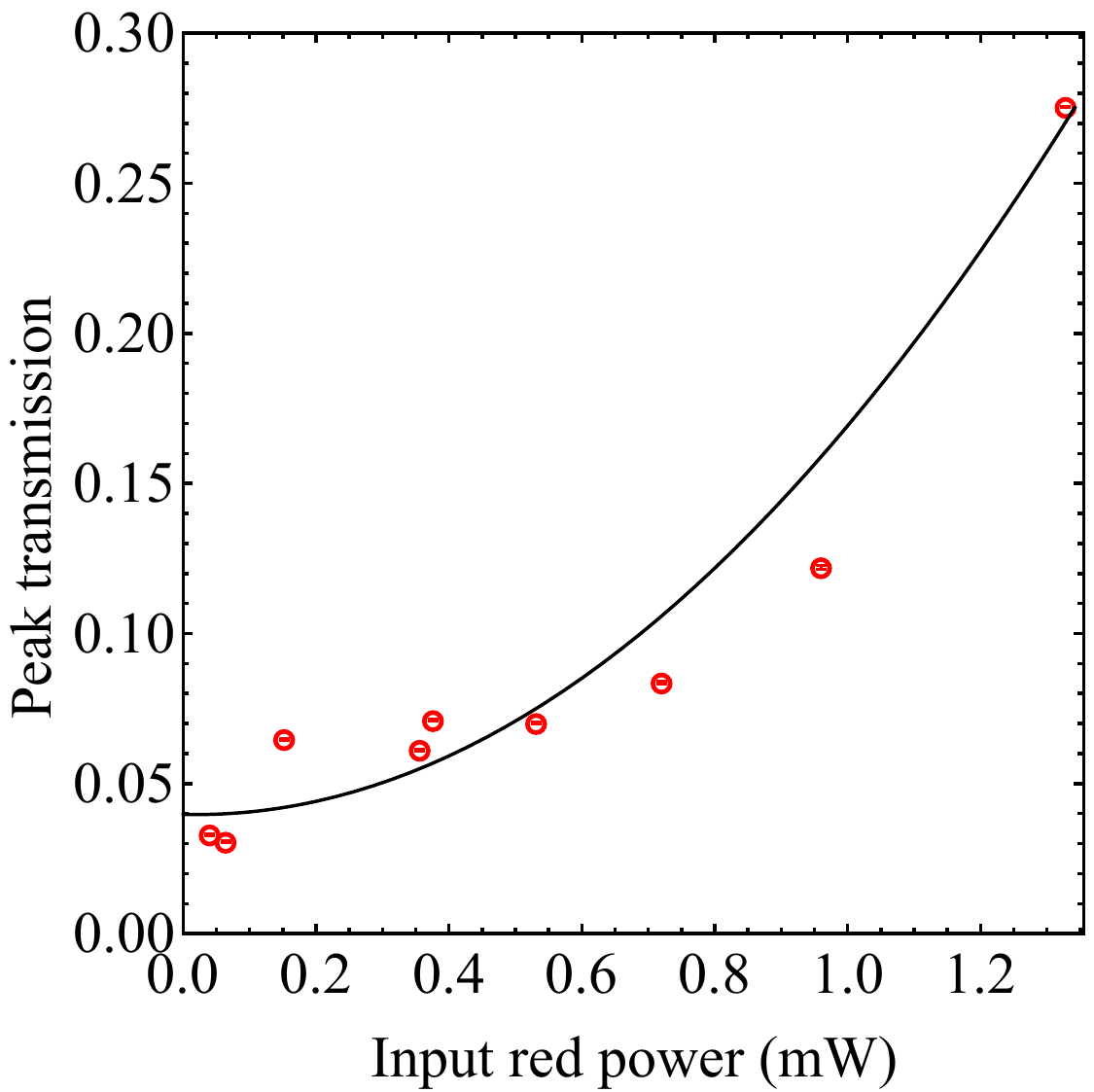}
    \caption{\label{fig:pktx} Peak transmission of zero-field $m_i{=}{\pm}1$ EIT resonances as a function of $P_{\rm red}$ along with quadratic fit.}
\end{figure}

We further make the assumption that the medium is optically thin. This may seem like an incorrect assumption, as $\alpha L$ ranges between $1.3\mbox{-}3$ (Fig.~\ref{fig:pktx}). However we believe the assumption is justified because most of the absorption is incoherent background absorption from NV centers which do not have $\Lambda$ transitions and/or belong to hyperfine states which are far detuned from two-photon resonance. Under this approximation, the EIT contrast and width can be calculated by evaluating the $\ket{E_x}$ population, $\rho_{\ket{E_x},\ket{E_x}}$, as a function of two-photon detuning, using Eq.~\eqref{eq:master}. In order to incorporate the effect of inhomogenous broadening of $\ket{E_x}$ levels, we sum over spectra from variable $\Delta_{0}$ over a range $-1$ to $1~{\rm GHz}$ (much larger than $\Omega_{s}$) \cite{SAN2006OPTEXS}.

In fitting the linewidth and contrast data [Fig.~\ref{fig:parameters}(a,b)] to spectra calculated in this way, we make the following assumptions. $\Omega_{s}$ is constrained to be $\propto\sqrt{P_{\rm red}}$. $\gamma_{gs,m_i}$ is allowed to be different for resonances associated with different nuclear spin projections ($m_i{=}0$ and $m_i{=}{\pm}1$). This is justified because the $m_i{=}0$ and $m_i{=}{\pm}1$ hyperfine levels are sensitive to different noise sources (electric versus magnetic field noise, respectively), as described in the main text and in \cite{DOL2011S}. $\gamma_1$ is constrained by the observed antihole width [Fig.~\ref{fig:EITspectra}(c)], $W$ (in Hz), as $\gamma_1{=}\Delta\nu_{\rm nat}\Omega_{0}^2/(2\pi W^2)$ \cite{KUZ2002S}. The value of $\gamma_1/(2\pi)$ is then of order a few kHz. This may seem to contradict recent results which predict $\gamma_1{=}1/T_1{\approx}2\pi\times0.01~{\rm Hz}$ \cite{JAR2012S}. However, those results were for NV centers in the absence of optical fields. We find that the in the presence of optical fields, the effects of, for example, spectral diffusion and interaction with other excited-state levels lead to a much larger effective $\gamma_1$ that describes our results.

The EIT contrast is sensitive to incoherent absorption from NV centers which do not exhibit $\Lambda$ transitions, hyperfine states which are off two-photon resonance, and the removal of NV$^{\mbox{-}}$ centers due to photo-ionization. The photo-ionization rate, in particular changes the power dependence, as it is quadratic in $P_{\rm red}$ \cite{ASL2013S}. Figure \ref{fig:pktx} shows the peak transmission of EIT spectra as a function of $P_{\rm red}$. The quadratic fit is further evidence that photo-ionization plays an important role in the EIT spectra. To account for these effects when fitting the simulated contrast to experimental data, we modify the calculated EIT contrast, $R_{\rm calc}$, as $R_{m_i}{=}a_{m_i} R_{\rm calc}/(1+b P_{\rm red}^2)$, where $a_{m_i}$ and $b$ are fitted parameters.

Under these approximations, we fit the model to the data in Fig.~\ref{fig:parameters}(a,b) and find the following fit parameters: $P_{\rm sat}{\equiv}\pi\Delta\nu_{\rm nat}^{2}P_{\rm red}/\Omega_{0}^2{=}2.4{\pm}1.1~{\rm mW}$, $\Omega_{1}/\Omega_{0}{=}0.08{\pm}0.02$, $\gamma_{gs,m_i{=}{\pm}1}/(2\pi){=}240{\pm}92~{\rm kHz}$, $\gamma_{gs,m_i{=}0}/(2\pi){=}99{\pm}30~{\rm kHz}$, $a_{m_i{=}{\pm}1}{=}1.0{\pm}0.1$, $a_{m_i{=}0}{=}0.3{\pm}0.1$, and $b{=}0.4{\pm}0.2~{\rm mW}^{-2}$.

\section{Modeling EIT resonances for higher temperature: 9-level model}

\begin{figure}
\centering
    \includegraphics[width=.4\textwidth]{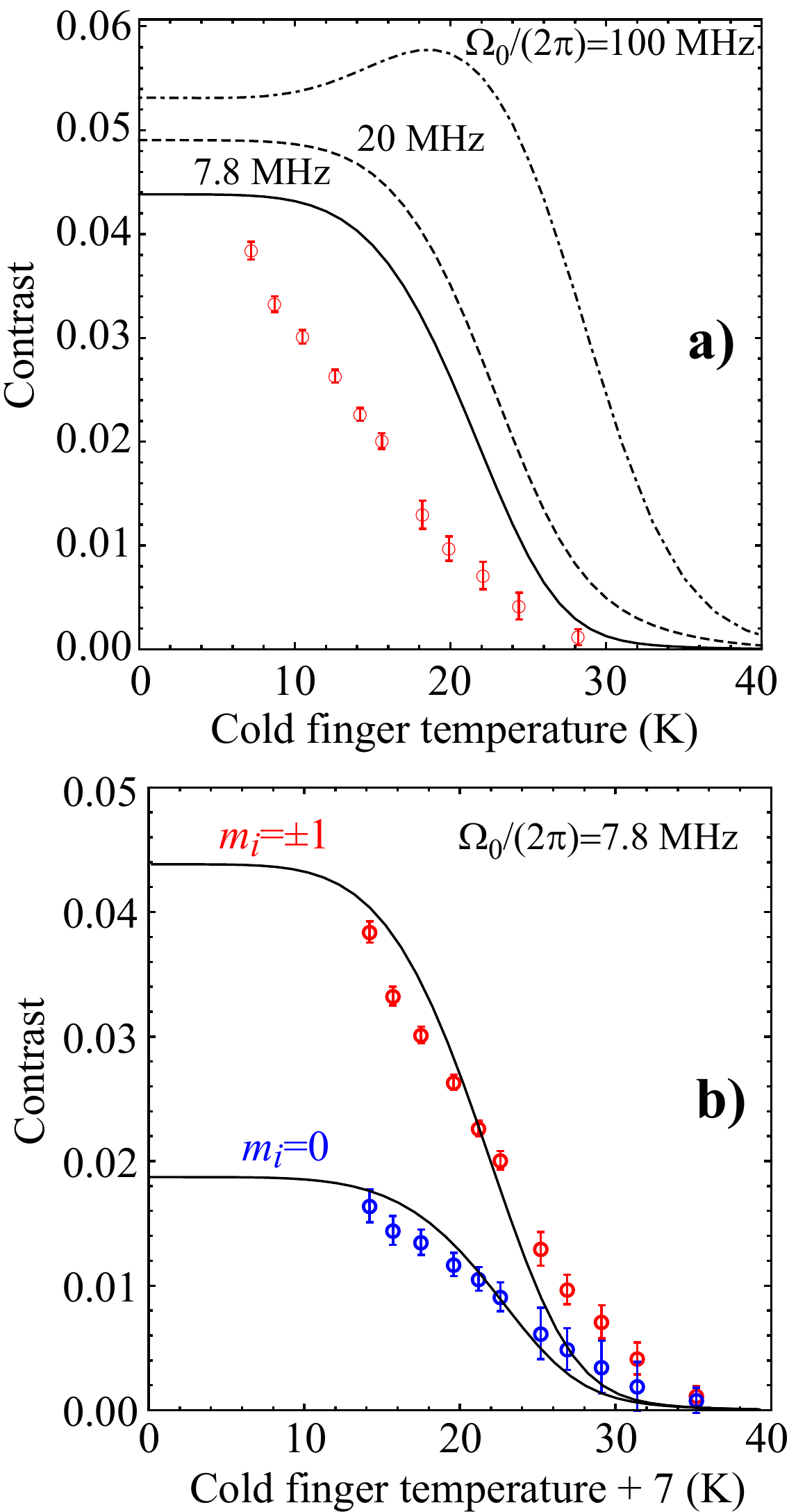}
    \caption{\label{fig:Tdep} (a) Temperature-dependence of the experimentally-determined contrast for the $m_i{=}{\pm}1$ zero-field EIT resonance ($\Omega_0{=}2\pi\times7.8{\pm}2.8~{\rm MHz}$). Also shown is the simulated contrast, using the 9-level model, for three values of $\Omega_0$. The simulated contrast is reduced by a factor of 13 to match the low-temperature experimental contrast. (b) Experimental contrast, shifted by 7 K, along with corresponding simulations. The simulated contrast is reduced by a factor of $13(35)$ for $m_i{=}{\pm}1(0)$ resonances.
}
\end{figure}

As briefly mentioned in the text, the EIT contrast falls off rapidly with increasing temperature. In our experiments, temperature is measured via a diode located on the copper ``cold finger'' on which the diamond is mounted. Figure \ref{fig:Tdep}(a) shows the EIT contrast of $m_i{=}{\pm}1$ resonances at $B{=}0$. The contrast becomes negligible for cold-finger temperatures $T{\gtrsim}30~{\rm K}$. In general, this phenomenon is due to phonon-induced orbital relaxation the excited state \cite{FU2009S}.

For the highest power used here, $P_{\rm red}{=}1.3~{\rm mW}$, the Rabi frequency can be calculated, using the 3-level-model fits described above, as $\Omega_0{=}2\pi\times7.8{\pm}2.8~{\rm MHz}$. As this Rabi frequency is much smaller than the energy spacing between levels in the excited state, the 3-level model can in principle be used to qualitatively describe the observed temperature dependence. This can be done by including temperature-dependent dephasing, $\Gamma_{es}(T)$ for both $\ket{s}\leftrightarrow{E_x}$ transitions, where $\Gamma_{es}(T)$ is determined using parameters in Ref. \cite{FU2009S}. In this model, the contrast is approximately proportional to $\Omega_0^2/[\Gamma_{es}(T)\gamma_{gs,m_i}]$. This would suggest that the deleterious effect on $\Gamma_{es}$, from increasing temperature, can be overcome by increasing $\Omega_0$. However, due to interactions with other excited-state levels, we find this is not the case.

A more general treatment considers interaction with all six excited state levels. For this 9-level model (all three ground-state sublevels and the six excited-state levels), we work in the Zeeman basis and include excited-state spin-orbit, spin-spin, and Stark-effect interactions \cite{DOH2011S}, pure ground-state spin dephasing and longitudinal relaxation (determined using 3-level model fits above), and excited-state homogenous broadening, $\Delta\nu_{\rm nat}{=}15~{\rm MHz}$. We also make the following assumptions:
\begin{itemize}
  \item inhomogenous broadening is neglected.
  \item the transverse strain is $15~{\rm GHz}$ (near a level ani-crossing).
  \item all optical coupling frequencies, $\Omega_0$, are equal.
  \item the laser frequencies are resonant with the lowest excited state.
  \item phonon-induced excited-state relaxation affects only the orbital portion of the wavefunction and preserves $m_s$.
  \item the contrast is calculated using the excited-state population (optically-thin medium).
\end{itemize}

Figure \ref{fig:Tdep}(a) shows the calculated contrast for three different values of $\Omega_0$. The simulated contrast is reduced by a factor of 13 to match the low-temperature experimental contrast. We find that increasing $\Omega_0$ to $2\pi\times100~{\rm MHz}$ only increases the contrast by a small amount, extending the operating temperature range by $\sim10$ degrees. At higher values of $\Omega_0$ the EIT lineshapes are severely distorted due to interaction with other excited states. Moreover, for $\Omega_0>2\pi\times100~{\rm MHz}$, no significant improvement in contrast at any temperature is observed. Note that in our experiments, the maximum Rabi frequency was $\Omega_0{\approx}2\pi\times8~{\rm MHz}$. At these low excitation rates, both the 3-level and 9-level models predict that the contrast is still increasing with increasing power, as observed in Fig.~\ref{fig:parameters}(b).

We also find that better quantitative agreement between experiment and theory can be obtained if we assume that the local temperature in the beam path is $7~{\rm K}$ greater than what is measured at the cold finger. Figure \ref{fig:Tdep}(b) shows the shifted experimental values along with their corresponding theoretical curves. The change in local temperature may be due to a combination of substantial optical absorption (${\sim}10~{\rm mW}$, time-averaged) and relatively poor thermal contact.

\section{Photon-shot-noise limited sensitivity as a function of $B$}
\label{sec:varyB}
Equation \eqref{eq:Bmin} in the main text describes the photon-shot-noise-limited minimum detectable field, $\delta B_{min}$, under the assumption that $|dS/dB|{=}g\mu_B\cos{\theta}R_1/\Delta\nu_{\rm eit}$. Here the contrast is $R_1{=}2A_1/(A_1+C)$ for $B{=}0$, but it is about half, $A/(A+C)$, when a sufficiently large bias field lifts the $m_s{=}{\pm}1$ degeneracy. We define the higher-field slope as $\xi{\equiv} g\mu_B\cos{\theta}\frac{1}{\Delta\nu_{\rm eit}}\frac{A}{A+C}$.

More generally, the slope, $dS/dB$, depends on both the lineshape and the bias field. It can be expressed as:
\begin{equation}
\label{eq:beta}
|dS/dB|=q\beta(B)\xi.
\end{equation}
The factor of $q$ is a numerical constant of order unity that accounts for the lineshape; $q{=}3\sqrt{3}/4{=}1.290$ for a Lorentzian lineshape. The bias-field-dependent factor, $\beta(B)$, arises due to the overlap of $\Delta m_s{=}{\pm}1$ resonances near $B{=}0$. This factor is important because it can dictate which bias field is optimal for a given value of $\Delta\nu_{\rm eit}$.

\begin{figure}
\centering
    \includegraphics[width=.4\textwidth]{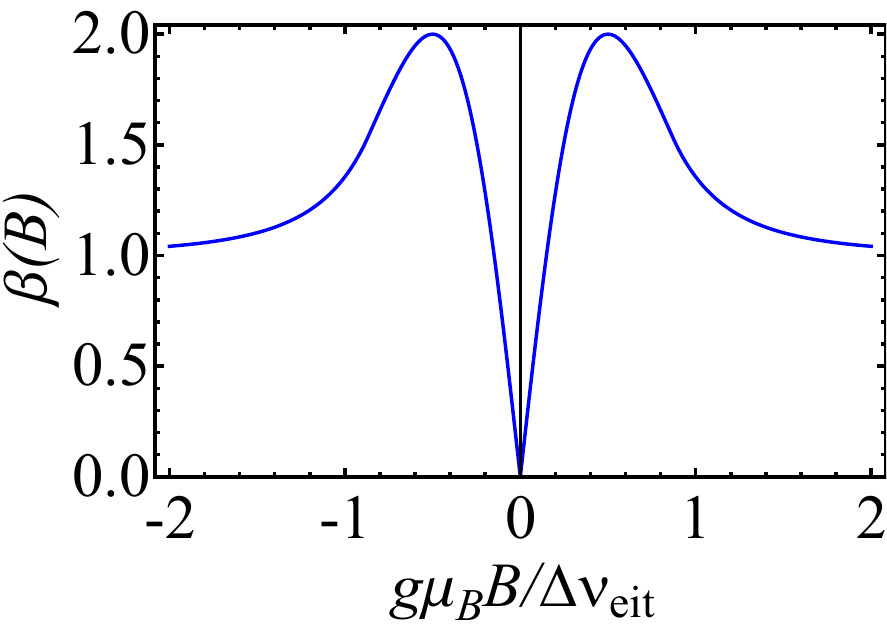}
    \caption{\label{fig:magopt} Magnetic response, $\beta(B)$, as a function of bias field along a [100] direction.}
\end{figure}

Figure \ref{fig:magopt} shows a plot of $\beta(B)$, assuming Lorentzian lineshapes. At exactly zero magnetic field, $\beta(0){=}0$ and the conventional lineshift magnetometry employed here is not possible. As $B$ increases, $\beta(B)$ rises sharply, reaching a maximum $\beta(B_{opt}){=}2$ at $g\mu_B|B_{opt}\cos{\theta}|{=}\Delta\nu_{\rm eit}/(2\sqrt{3})$. For large fields, satisfying $g\mu_B|B|{\gg}\Delta\nu_{\rm eit}/2$, $\beta(B)$ approaches unity.

The sharp dependence of $dS/dB$, and consequently $\delta B_{min}$, on bias field may have important implications for proposed fundamental physics experiments. For example, Ref.~\cite{BEC2011S} proposes to search for the electric dipole moment of neutrons in an ambient magnetic field $B{=}1~{\mu T}$.  Using typical experimental parameters in this work ($P{=}100~{\mu W}$, $A/(A+C){=}0.035$, $\Delta\nu_{\rm eit}{=}1~{\rm MHz}$, $\cos{\theta}{=}1/\sqrt{3}$, $q{=}3\sqrt{3}/4$), we find that, at $|B|{=}1~{\rm \mu T}$, the minimum detectable field would be $\delta B_{min}{\simeq}420~{\rm pT/\sqrt{Hz}}$. This is nearly an order of magnitude larger than the optimal sensitivity, $\delta B_{min}{\simeq}50~{\rm pT/\sqrt{Hz}}$ at $|B_{opt}|{=}18~{\rm \mu T}$.

However, improvement in sensitivity at low bias field ($g\mu_B|B|{\ll}\Delta\nu_{\rm eit}/2$) can be realized by decreasing $\Delta\nu_{\rm eit}$. This is because in this regime $\beta(B){\propto}1/\Delta\nu_{\rm eit}$, so consequently $\delta B_{min}{\propto}\Delta\nu_{\rm eit}^2$ [Eq.~\eqref{eq:beta}]. If instead $\Delta\nu_{\rm eit}{=}0.25~{\rm MHz}$, and the contrast does not change, then $\delta B_{min}{\simeq}27~{\rm pT/\sqrt{Hz}}$ at $|B|{=}1~{\rm \mu T}$, a $16$-fold improvement. Narrowing of the EIT resonances could be accomplished by reducing $P_{\rm red}$ or using isotopically pure diamond. Optimal sensitivity could also be realized by actively applying an appropriate bias field.

We note that, as discussed in the text, electric-field sensitivity is realized only under sufficiently low bias magnetic field. For $m_i{=}0$ resonances, the sensitivity falls off by a factor of ${\sim}\sqrt{2}$ under a bias field, $|B_{max}|{\equiv} d_{gs_{\perp}}|E_{\perp}|/(g\mu_B\cos{\theta})$. For the effective field $|E_{\perp}|{\approx}15~{\rm kV/cm}$ present in our sample, this value is $|B_{max}|{\approx}15~{\rm \mu T}$. Fortunately, this bias field is comparable to the bias field where $\delta B_{min}$ is optimal, $|B_{opt}|{=}18~{\rm \mu T}$, ensuring that high-sensitivity electrometry and magnetometry can be simultaneously realized.

\section{Comment on E-field sensing}
The electric-field sensitivity quoted here is for fields inside of diamond. If the goal of the electrometer is to sense external charges/fields, one needs to account for dielectric screening. In most geometries, the electric field measured in vacuum will be larger than the field in diamond by a factor of order the dielectric constant, $5.7$.

\section{Magnetometer linearity}
In the present implementation, the range over which the magnetometer output remains linear, $\Delta B_{\rm lin}$, is constrained by the EIT linewidth. In particular, using the microwave frequency-modulation technique, $\Delta B_{\rm lin}$ is always smaller than the modulation depth, $\Delta\nu_{\rm mod}$. $\Delta\nu_{\rm mod}$ is chosen to optimize the magnetometer scale factor (conversion of lock-in output to magnetic field units), $\Delta\nu_{\rm mod}\approx\Delta\nu_{\rm eit}/\sqrt{3}$ (see Fig. \ref{fig:magopt} and related discussion).

\begin{figure}
\centering
    \includegraphics[width=.4\textwidth]{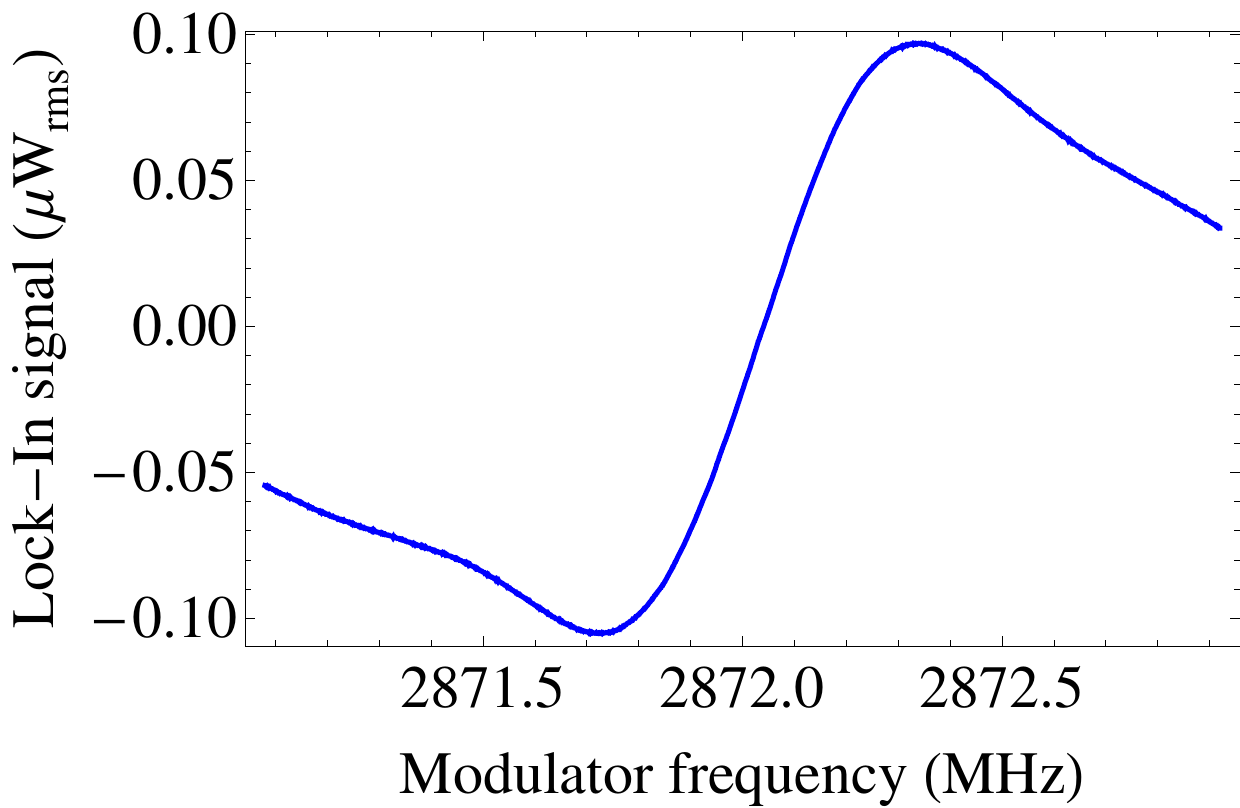}
    \caption{\label{fig:lockinsweep} (a) Lock-in signal as a function of microwave carrier frequency.}
\end{figure}

Figure \ref{fig:lockinsweep} shows the lock-in signal as a function of carrier frequency using the experimentally-optimized modulation depth, $\Delta\nu_{\rm mod}=300~{\rm kHz_{pp}}$. From the figure, we see that the scale factor remains linear in a range of $\Delta\nu_{\rm lin}\approx100~{\rm kHz}$, corresponding to $\Delta B_{\rm lin}=\Delta\nu_{\rm lin}/(g\mu_B\cos{\theta})\approx20~{\rm \mu T}$.

\begin{figure}
\centering
    \includegraphics[width=.48\textwidth]{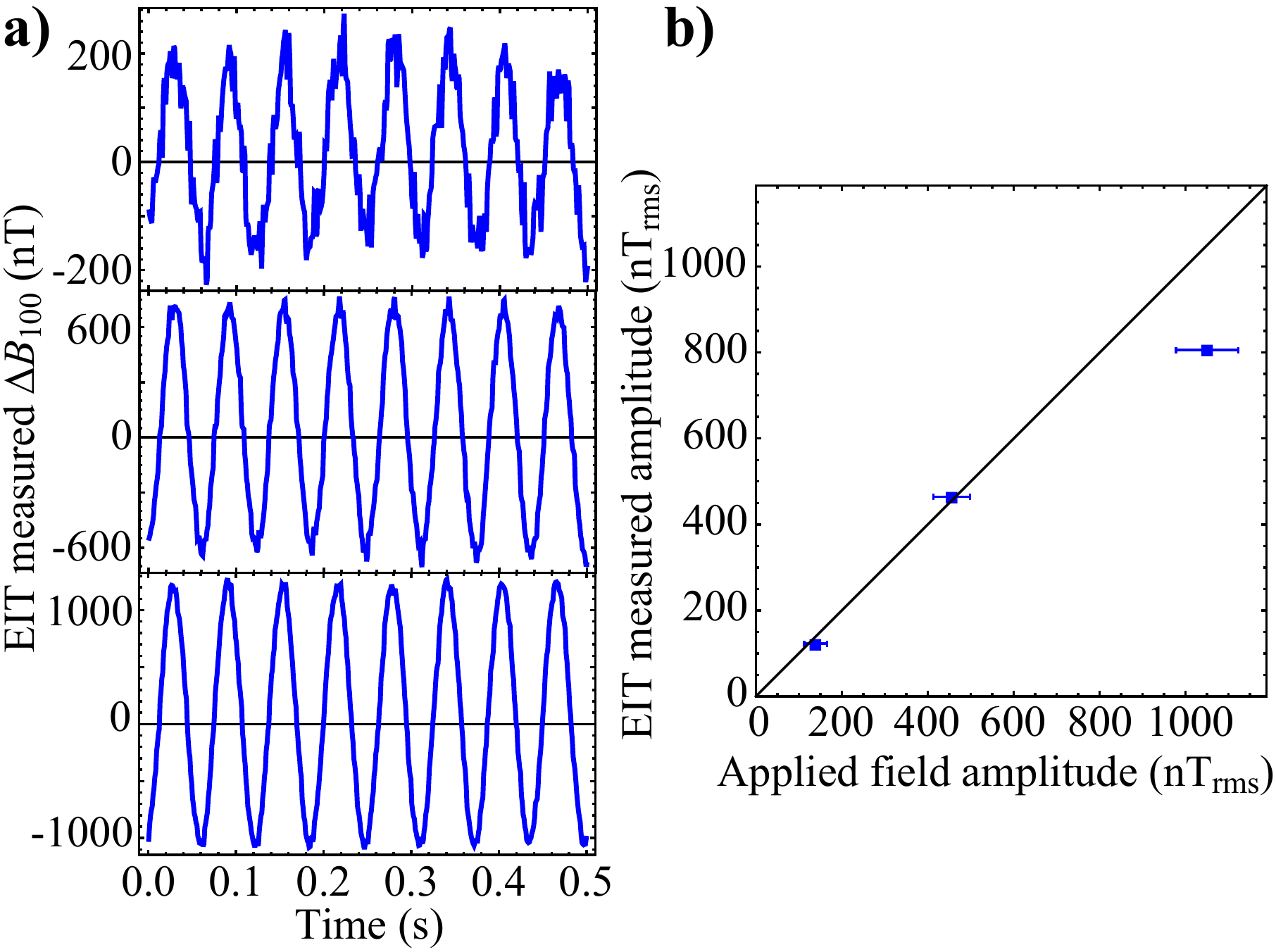}
    \caption{\label{fig:linearity} (a) EIT magnetometer response to 16-Hz applied fields of three different amplitudes. (b) Measured field amplitude as a function of the expected applied field amplitude. The solid line corresponds to perfect linearity (slope = 1) with no offset.}
\end{figure}

To verify the linear response, we applied 16-Hz sinusoidal current modulations to a coil with different amplitudes, and measured the resulting EIT magnetometer output. The conversion of applied current to applied $B_{100}$-field was extrapolated by measuring the shifts in EIT spectra to large changes in current (as in Fig. \ref{fig:EITspectra}(c) in the main text). Figure \ref{fig:linearity}(a) shows the lock-in response to each applied field. The measured amplitude is compared to the separately-calibrated input field amplitude in Fig. \ref{fig:linearity}(b). As expected, the magnetometer remains linear for the two smallest applied fields, corresponding to a range $\Delta B_{\rm lin}\approx1~{\rm \mu T}$. For the largest amplitude ($1.1~{\rm \mu T_{rms}}$) there was some deviation from the expected response; this may be due to, for example, a mismatch of the lock-in reference phase.

\section{Magnetometer scale-factor drift}
\begin{figure}
\centering
    \includegraphics[width=.48\textwidth]{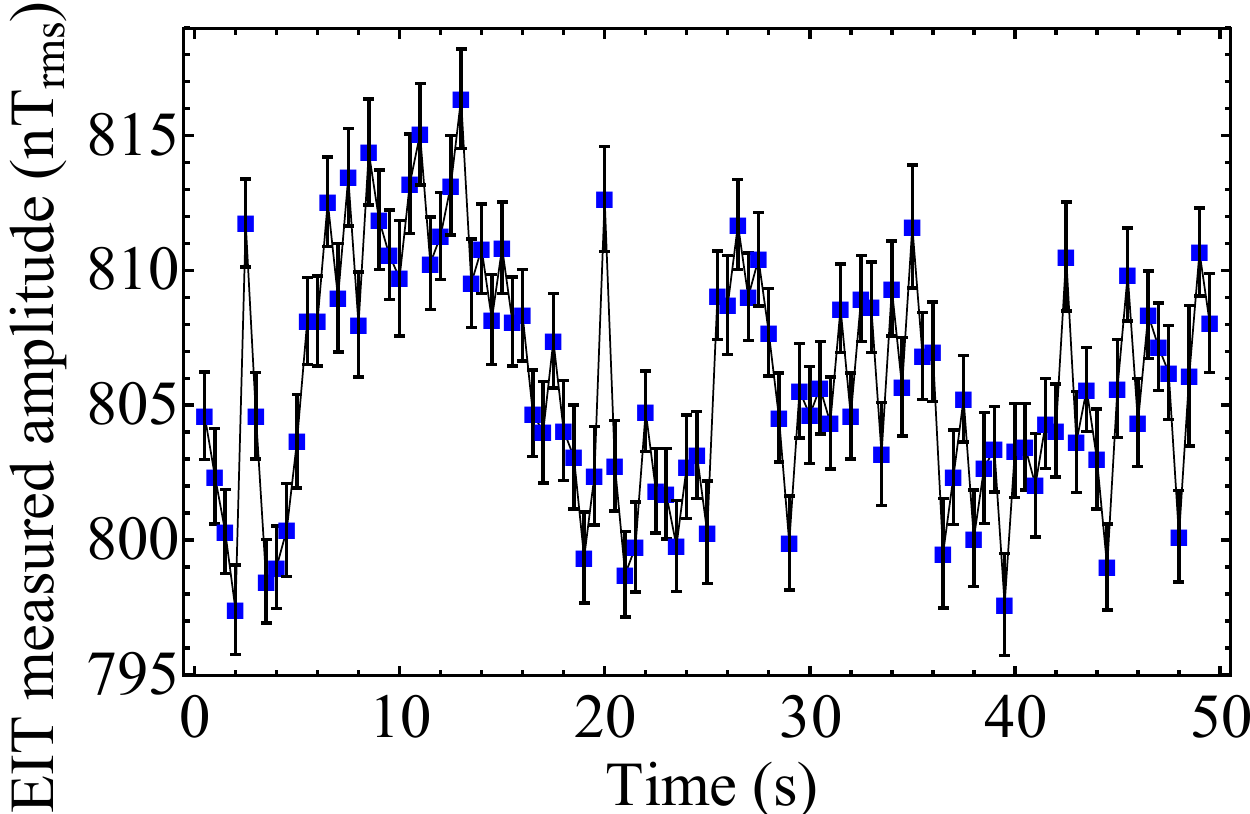}
    \caption{\label{fig:scalefactor} (a) Fitted amplitude of a 16-Hz applied field over the course of time.}
\end{figure}

The long-term accuracy of the sensor relies on the scale factor remaining constant over time. To test the scale-factor drift, we applied a 16-Hz sinusoidal field for $50~{\rm s}$, and monitored the measured amplitude over time by fitting a sine curve to $0.5$-s intervals. Figure \ref{fig:scalefactor} shows the results of the measurement. No clear long-term drift of the measured amplitude is observed, so we conservatively bound the scale-factor drift at $\lesssim1\%/{\rm min}$. Note that any residual drift can be corrected for by, for example, periodically monitoring the lock-in response curve or adding an additional EOM frequency modulation with a different depth and frequency.

\section{Magnetometer bandwidth}
The fundamental limit on our EIT magnetometer bandwidth is set by the polarization rate of NV centers into the dark state, $\sim\Delta\nu_{\rm eit}$. In the present implementation, the bandwidth was limited by the electronic filtering of the lock-in amplifier. The amplifier used a series of four 6 dB/octave low-pass filters, resulting in an attenuation to signals at frequency, $f$, of $(1+2\pi\tau_L f)^{-2}$. In order to avoid artifacts introduced by the 4 kHz repetition rate of the optical pulses, a filter time constant, $\tau_L=1~{\rm ms}$, was used.

\begin{figure}
\centering
    \includegraphics[width=.4\textwidth]{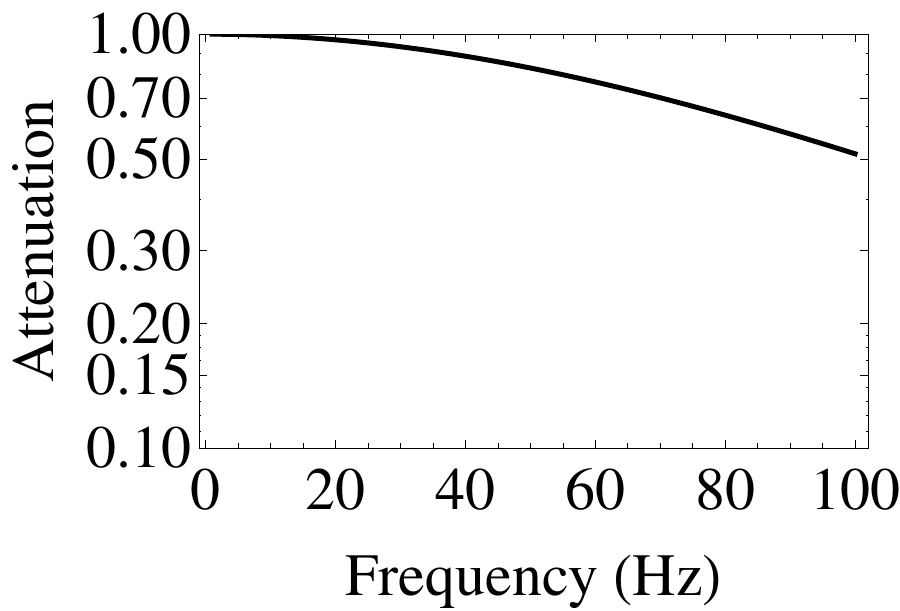}
    \caption{\label{fig:bandwidth} (a) Attenuation of the EIT magnetometer signal due to the electronic filtering used here.}
\end{figure}

Figure \ref{fig:bandwidth} shows the electronic attenuation as a function of frequency. We tested the magnetometer frequency response by varying the frequency of a $\sim130~{\rm nT_{rms}}$ sinusoidal field modulation, and found excellent agreement with the response plotted in Fig. \ref{fig:bandwidth}. This effect is also responsible for the slight downward slope of the noise floor at frequencies $\gtrsim10~{\rm Hz}$ in Fig. \ref{fig:EITmag}(c) of the main text, as those data were not corrected for the small frequency-dependent attenuation.

\section{Magnetometer noise}
From Figures \ref{fig:EITmag}(c-d) in the main text, it is evident that there is low-frequency noise near 1 Hz. We investigated the origin of this noise by recording noise spectra under different conditions, including:
\begin{enumerate}
  \item the photodetector was unplugged, and we recorded the lock-in output [Fig.~\ref{fig:EITmag}(c)].
  \item $\nu_{\rm EOM}$ was detuned $-3~{\rm MHz}$ from two-photon resonance and we recorded the lock-in signal with the photodetector connected.
  \item the signal from a commercial fluxgate magnetometer, placed close to the position of the diamond, was recorded.
\end{enumerate}
However, in all these cases, the noise was absent. It is worth mentioning that the fluxgate magnetometer also picked up magnetic noise at $60~{\rm Hz}$ of a similar amplitude to the noise peak recorded by the EIT magnetometer [Fig.~\ref{fig:EITmag}(c) in the main text]. However the noise floor near $1~{\rm Hz}$ was ${\lesssim}1~{\rm nT/\sqrt{Hz}}$.

We estimate that AC Stark shifts of order $\Omega_0^2/(16\pi^2\Delta_{e}){\approx}10~{\rm kHz}$, are present under our typical operating conditions. This was confirmed by EIT spectra obtained from the nine-level model described above. Consequently fluctuations in $P_{\rm red}$ of $1\%$ lead to variations in magnetometer output of order a few nT. Such fluctuations are consistent with the magnitude of the noise. We note that fluctuations in power also cause the overall transmission to change, but this generally happens slow compared to our modulation frequency (89 kHz) so it does not show up in the magnetometer signal (as evidenced by the lack of the noise when $\nu_{\rm EOM}$ was detuned $-3~{\rm MHz}$). The AC-Stark fluctuations may be caused by vibrations due to turbulent cryostat flow and/or vibrations of the transfer line connecting the cryostat to the liquid helium dewar. Future low-frequency applications may benefit from active power stabilization or post-processing the magnetometer signal based on monitoring the power.


%

\end{document}